\documentclass[aps,prd,twocolumn,floatfix,nofootinbib,superscriptaddress,
showpacs,amssymb]{revtex4}

\usepackage{bm}
\usepackage{graphicx}

\def\bea{\begin{eqnarray}}
\def\eea{\end{eqnarray}}
\def\beq{\begin{equation}}
\def\eeq{\end{equation}}

\begin{document}

\title{Epicyclic frequencies for rotating strange quark stars:\\
the importance of stellar oblateness}

\newcommand{\UNI}{Institute of Astronomy, University of Zielona G\'ora, Lubuska 2, 65-265, Zielona G\'ora,  Poland}
\newcommand{\CAMK}{Nicolaus Copernicus Astronomical Center, Bartycka 18, 00-716 Warsaw, Poland}
\newcommand{\GREECE}{Departament of Physics, Aristotle University of Thessaloniki, Thessaloniki, 54124 Greece}
\author{Dorota Gondek-Rosi\'nska}\email{dorota@astro.ia.uz.zgora.pl}\affiliation{\UNI}
\author{W\l odek Klu\'zniak}\email{wlodek@camk.edu.pl}\affiliation{\CAMK}
\author{Nikolaos Stergioulas}\affiliation{\GREECE}
\author{Mateusz Wi\'sniewicz}\affiliation{\UNI}


\begin{abstract}

Kilohertz QPOs can be used as a probe of the inner regions of
accretion disks in compact stars and hence also of the properties of
the central object. Most models of kHz QPOs involve epicyclic frequencies
to explain their origin.  We compute the epicyclic frequencies
of nearly circular orbits around rotating strange quark stars. The MIT
bag model is used to model the equation of state of quark matter and
the uniformly rotating stellar configurations are computed in full
general relativity. The vertical epicyclic frequency and the related
nodal precession rate of inclined orbits are very sensitive to the
oblateness of the rotating star. For slowly rotating stellar models of
 moderate and high mass strange stars, the sense of the nodal precession
changes at a certain rotation rate. At lower stellar rotation rates
the orbital nodal precession is prograde, as it is in the Kerr metric,
while at higher rotation rates the precession is retrograde, as it is
for Maclaurin spheroids. Thus, qualitatively, the orbits around
rapidly rotating strange quark stars are affected more strongly by the
effects of stellar oblateness than by the effects of general
relativity. We show that epicyclic and orbital frequencies calculated
numerically for small mass strange stars are in very good agreement
with analytical formulae for Maclaurin spheroids.

\end{abstract}


\maketitle

\section{Introduction}
\label{intro}

Strange quark stars (SQS) are considered as a possible alternative to
neutron stars as compact objects (see, e.g., \cite{Weber99} for a
review). The possibility of the existence of quark matter was first
recognized in  the early seventies. Bodmer \cite{Bodmer71}
 remarked that matter
consisting of deconfined up, down and strange quarks could be the
absolute ground state of matter at zero pressure and temperature.  If
this is true, then macroscopic stellar-mass objects made of such
matter, i.e., quark stars (also called  ``strange stars") could in
principle exist \cite{Witten84}. Typically, 
strange stars \cite{Alcock86,Haens86} are modeled
 with an equation of state (EOS)
 based on the phenomenological MIT-bag model of quark matter in which
quark confinement is described by an energy term proportional to the
volume \cite{Fahri84}.

It was shown \cite{Gonde03} that a strange star
described by the standard MIT bag model can be accelerated to high
rotation rates in low-mass X-ray binaries (LMXBs) taking into account
both a reasonable value of mass of the strange quark, and secular
instabilities, such as the viscosity-driven instability and the
$r$-mode instability.  Therefore, strange stars in LMXBs could rotate
rapidly (with spin frequency $> 400\,$Hz).  This provides the
astrophysical motivation for computing models of rapidly rotating strange
stars.

General relativity (GR) predicts the existence of the marginally stable
orbit, within
which no stable circular motion is possible \cite{Kapla49}, and all
models of accretion disks around black holes take this into account;
e.g. \cite{Shaku73,Sadow11}.  In the case of neutron stars,
 the marginally
stable orbit may be separated from the stellar surface by a gap, whose
size depends on the equation of state of dense matter and the
spin of the neutron star, as well as on its mass
 \cite{Kluzn85,Cook94}.  Whether or not the accreting
fluid attains that orbit depends also on the value of the stellar
magnetic field \cite{Kluzn90}. Similar
considerations apply to quark stars \cite{Gonde01}.

The marginally stable orbit is often called the innermost stable
circular orbit (ISCO). While in the Kerr geometry 
\cite{Bardeen72} this can lead to no
misunderstandings, in general an ISCO cannot be identified with the
marginally stable orbit.  In some metrics a marginally stable orbit
may be the outermost (in a certain radius range) stable circular orbit
\cite{Pugli11,Vieira13},
 while in the Newtonian gravity of a $1/r$ potential all
circular orbits are stable, and the innermost one is simply the one
grazing the surface of the spherical gravitating body. In this paper
we will mostly avoid using the term ISCO in the context of quark stars,
where the term ISCO is unambiguous only when there is a gap between
the marginally stable orbit and the stellar surface.
 
Whenever the marginally stable orbit is present around 
a neutron star or a quark star, its frequency  is an
upper bound on the frequency of stable orbital motion of a test
particle.  In addition to the
orbital frequencies, epicyclic frequencies are of great interest in
the discussion of accretion disks in GR.  Indeed, the Rayleigh
criterion for stability of circular motion is that $\nu_r^2>0$. Thus,
the radial epicyclic frequency, $\nu_r$, goes to zero at the
marginally stable orbit, and therefore must have a maximum at a
somewhat larger  radius.\footnote{In Schwarzschild geometry the ISCO
  is at $r=6M(G/c^2)$, while the maximum of $\nu_r$ is attained at
  $r=8M(G/c^2)$.}  The presence of a maximum in $\nu_r$ allows mode
trapping of $g$-modes of disk oscillations, whose eigenfrequency is
somewhat lower than the maximum value of $\nu_r$ \cite{Kato80,Nowak92},
 while the vertical epicyclic frequency is related to a
generalization of the Lense-Thirring precession, the so called
$c$-mode, whose eigenfrequency is approximately equal to the
difference between the orbital and the vertical epicyclic frequencies
\cite{Silbe01}.  Such modes may have
been detected in LMXBs as the celebrated kHz QPOs
 (see e.g., \cite{Klis00} for a review of QPOs).

In Newtonian gravity, all circular orbits around spherically symmetric
objects are stable. However, the marginally stable orbit may be present
around {\sl rapidly rotating} Newtonian stars \cite{KluznBG01,Zduni01}. 
Indeed, for rapidly rotating Maclaurin spheroids, the ISCO is well
outside the surface of this figure of equilibrium \cite{Amste02}.

In this paper we report on numerical calculations in general
relativity of epicyclic frequencies for rotating strange stars
with astrophysically relevant masses of $1.4 M_{\odot}$ and
$1.96 M_{\odot}$. We use an up-to-date version of the RNS code \cite{StergF95}.
The motivation for the study is explained in \S~\ref{aims},
and the implications of our findings are sketched in \S~\ref{astro}.

\section{Aims of the study}
\label{aims}

The primary purpose of this work is to understand the influence of
rotation-induced oblateness on the orbital and epicyclic frequencies
in rapidly rotating relativistic stars. In particular, we would like
to clarify the origin of the departures of these frequencies
from their Kerr values. Astrophysically this is interesting, for instance
in the context of similarities and differences between the kHz QPOs
in LMXB neutron stars and their counterparts in black hole X-ray sources.
Previously, in the context of
neutron star QPOs, it had been expected that
frame-dragging effects dominate those of oblateness \cite{Kluzn98,Morsink99},
as they do in the Hartle-Thorne metric \cite{Hartle68}.

At the same time, we would like to validate the RNS code in the
exacting regime of the low-mass Newtonian limit. 
In fact we are in the fortunate position of having exact analytic
formulae \cite{Gonde13} against which numerical values of the frequencies
can be compared. Previously, the only test available for the epicyclic
frequency module was given by the position of the ISCO, for which the
RNS numerical value could
be compared with the known formulae for the
Schwarzschild and Hartle-Thorne metrics, or with the numerical values
obtained in the LORENE code \cite{Lorene}
 (however, LORENE does not as yet allow
computation of the epicyclic frequencies).

For these two reasons we have chosen to study quark star models.
Quark matter, being self-bound (if it exists), admits stellar
models of arbitrarily low mass, while neutron stars being bound
by gravity are intrinsically relativistic objects (in the sense
of GR) and are  unstable at low masses.
We have performed calculations
for extremely low-mass models of gravitational mass $M = 0.001M_\odot$
and $0.01M_\odot$ to check the accuracy of our relativistic
code in the Newtonian limit. Our numerical calculations
of the orbital and epicyclic frequencies around low mass stars
agree very well with the analytical calculations
for Maclaurin spheroids \cite{Amste02,Gonde13}, thus validating the code in
the low-mass limit as well as its epicyclic frequency module in general.

 Further, quark stars have higher density than neutron stars and
(apart from a possible thin baryonic crust) are devoid of low density
envelopes. Hence, the effects of stellar oblateness related to rotation
are more pronounced than in neutron stars. This allows a fairly clear
identification of the oblateness contribution, as we will see
in the results of our calculations of models with astrophysically
relevant masses.

Of course, having at hand results for rapidly rotating quark-star models,
we can try to speculate on the implications for observed phenomena,
such as QPOs, and we do so briefly in \S~\ref{astro}.
It has long been recognized that a secure identification of quark
stars would have profound implications for our view of the stability
of hadronic matter \cite{Bodmer71, Witten84},
 and it is in this spirit that predicted observational
signatures of quark stars have been studied; e.g., \cite{Madsen99,Sterg99}.

\section{Equation of state of strange stars}
\label{eos}

We perform all numerical calculations of quark stars
in the framework of the MIT
bag model, within which the quark matter is composed of massless up
 and down quarks, massive strange quarks, and electrons.
There are three physical quantities
describing the model: the mass of the strange quarks, $m_{\rm s}$, the
bag constant, $B$, and the strength of the QCD coupling constant
$\alpha$.  
We use a simple MIT bag model where $m_{\rm s}=0$,
$\alpha=0$ and $B=60\ {\rm MeV/fm^3}$, with the equation of state given by 
$$P=a(\rho-\rho_0)c^2, \eqno (1)$$ 
where $P$ is the pressure, $\rho$
the mass-energy density, and $c$ is the speed of light. 
Both $\rho_0$ and $a$ are functions of the physical
constants $B$, $m_{\rm s}$ and $\alpha$. In our case $a=1/3$
and $\rho_0=4.2785 \times 10^{14} {\rm g/cm^3}$.   Essentially, Eq. (1)
corresponds to self-bound matter with density $\rho_0$ at zero
pressure and with a fixed sound velocity ($\sqrt{a}\,c$) at all
pressures.  For a fixed value of $a$, all stellar parameters are
subject to scaling relations with appropriate powers of $\rho_0$.
For instance,
 $f\propto \rho_0^{1/2}$, where $f$ denotes either of the
rotational or orbital frequencies, and $M, R \propto \rho_0^{-1/2}$ (e.g.,
\cite{Witten84,ZduniBKHG00}).

\begin{figure}
\begin{center}
\includegraphics[angle=-90,width=0.52\textwidth]{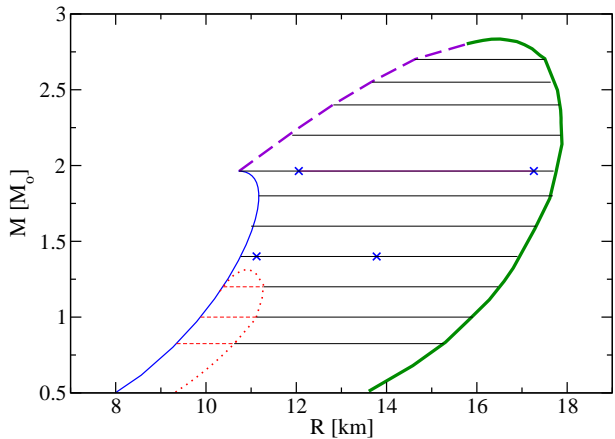}
\caption{Gravitational mass versus equatorial radius for static (thin
  blue line) and uniformly rotating strange quark stars described by
  the MIT bag model. Horizontal lines correspond to sequences with a
  constant gravitational mass. The radius increases with rotation rate
  until the mass-shedding limit is reached (thick green line). Both
  normal and supramassive sequences are shown. Thick solid violet
  long-dashed line corresponds to configurations marginally stable to
  axisymmetric perturbations. The crosses represent models
  considered in detail in the paper: two with gravitational mass $1.4
  M_\odot$, rotating at frequencies $600\,$Hz and $1165\,$Hz (from
  left to right), and two with $1.964 M_\odot$ at $910\,$Hz and
  $1252\,$Hz. }
\label{fig1}                                 
\end{center}                                
\end{figure}
\begin{figure}
\begin{center}
\includegraphics[angle=-90,clip,width=0.52\textwidth]{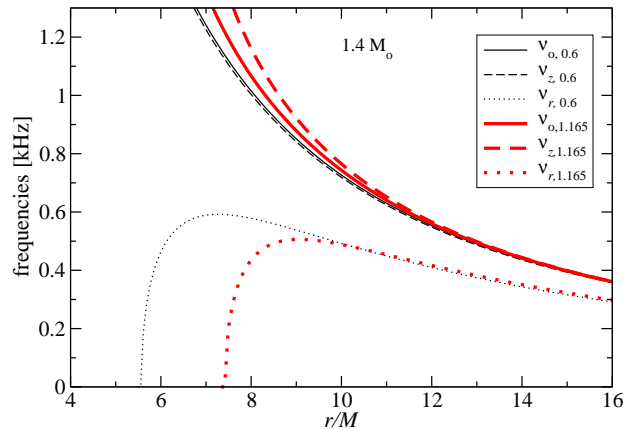}
\caption{Orbital and epicyclic frequencies versus
  radius (scaled with gravitational stellar mass $M$) for numerical models of
  an $M=1.4 M_\odot$ uniformly rotating strange quark star
  rotating at a fixed frequency,
  600 Hz (thin black lines) and  1165 Hz (thick red lines). 
  From top to bottom: vertical epicyclic frequency (dashed
  lines), orbital frequency (solid lines), radial epicyclic frequency
  (dotted lines).}
\label{velocity14}
\end{center}
\end{figure}

\section{Rotating strange quark stars}

\subsection{Numerical models}
\label{num}
We have calculated axisymmetric models of rotating strange quark stars
and their exterior metrics using a highly accurate relativistic code,
RNS (\cite{StergF95}, see \cite{Sterg03} for a
description). In this code the equilibrium models are obtained
following the KEH method \cite{Kamat89}, with extensions by
\cite{Cook94}, in which the field equations are
converted to integral equations using appropriate Green's
functions. The code was applied to perform calculations of epicyclic
frequencies for neutron stars in \cite{Kluzn04}, and the
extended RNS code was tested against selected results published in
\cite{Marko00}.

We have computed the metric outside uniformly rotating quark stars of
masses and rotation rates that may be typical of compact stars which
have been spun up through accretion in LMXBs.  In this paper we will
consider in detail the orbital and epicyclic frequencies of  test
particles for four models of quark stars with masses and rotation
rates likely to occur in LMXBs, if there are LMXBs containing quark
stars. These are two models of stars with gravitational mass
$M=1.4M_\odot$,  rotating at frequencies  $\nu_* = 600$ Hz and at
 $\nu_*=1165$ Hz, and two models with $M=1.964 M_\odot$ rotating at
$\nu_* =910$ Hz and $\nu_* =1252$ Hz.  These models are indicated
by crosses in Fig.~\ref{fig1} (from left to right and bottom to top). 

The general relationship of gravitational mass to the equatorial
radius for uniformly rotating SQS described by the MIT bag model is
shown in Fig.~\ref{fig1}.  Each horizontal (thin solid or
short-dashed) line corresponds to a sequence with constant
gravitational mass.  For each sequence the largest radius is attained
for the equatorial mass-shedding models (thick green solid line),
while the smallest radius model is either the static model 
(leftmost, thin, blue solid curve)
 or a rotating configuration marginally stable to the
axisymmetric perturbations (thick violet long-dashed line).  Models
with higher baryon mass than the baryon mass of static neutron star
with maximum gravitational mass are called supramassive.  The angular
momentum increases along each sequence from $J=0$ for static
configurations, or $J_ {\rm min}>0$ on the axisymmetric stability
limit, to $J_{\rm max}$ for configurations at the mass-shedding limit.
Mass shedding occurs when the angular velocity of the star reaches the
angular velocity of a particle in a circular Keplerian orbit at the equator
(for SQS this is an unstable orbit).

Sequences shown with short-dashed red lines correspond to rotating
models for which the marginally stable orbit does not exist.  The
limiting configurations for which the radius of the marginally stable
orbit coincides with the stellar equatorial radius, i.e., for which
$r_{\rm ms}=R_{\rm eq}$, are shown by the thin red dotted line in
Fig.~\ref{fig1}. This line separates the models with and without a
marginally stable orbit.  It is evident that the marginally stable
orbit exists around SQS for a broad range of masses and rotation rates
and is absent only in slowly rotating, low-mass strange quark  stars.

\subsection{Orbital and epicyclic frequencies}
\label{freq}
If one recalls that the radius of the marginally stable orbit
decreases with increasing black hole spin in the Kerr metric, and that
the stellar radius increases with increasing stellar angular momentum,
it may seem counter-intuitive that the gap between the marginally
stable orbit and the
stellar surface disappears for low stellar rotation rates, while it is
present for rapidly rotating models.  Of course, the space-time of a
rotating star is not described by the Kerr metric, and the complex
behavior of the marginally stable orbit, and more generally of the
epicyclic frequencies for rapidly rotating compact stars, is the focus
of this paper.


\begin{figure}
\begin{center}
\includegraphics[angle=-90,clip,width=0.52\textwidth]{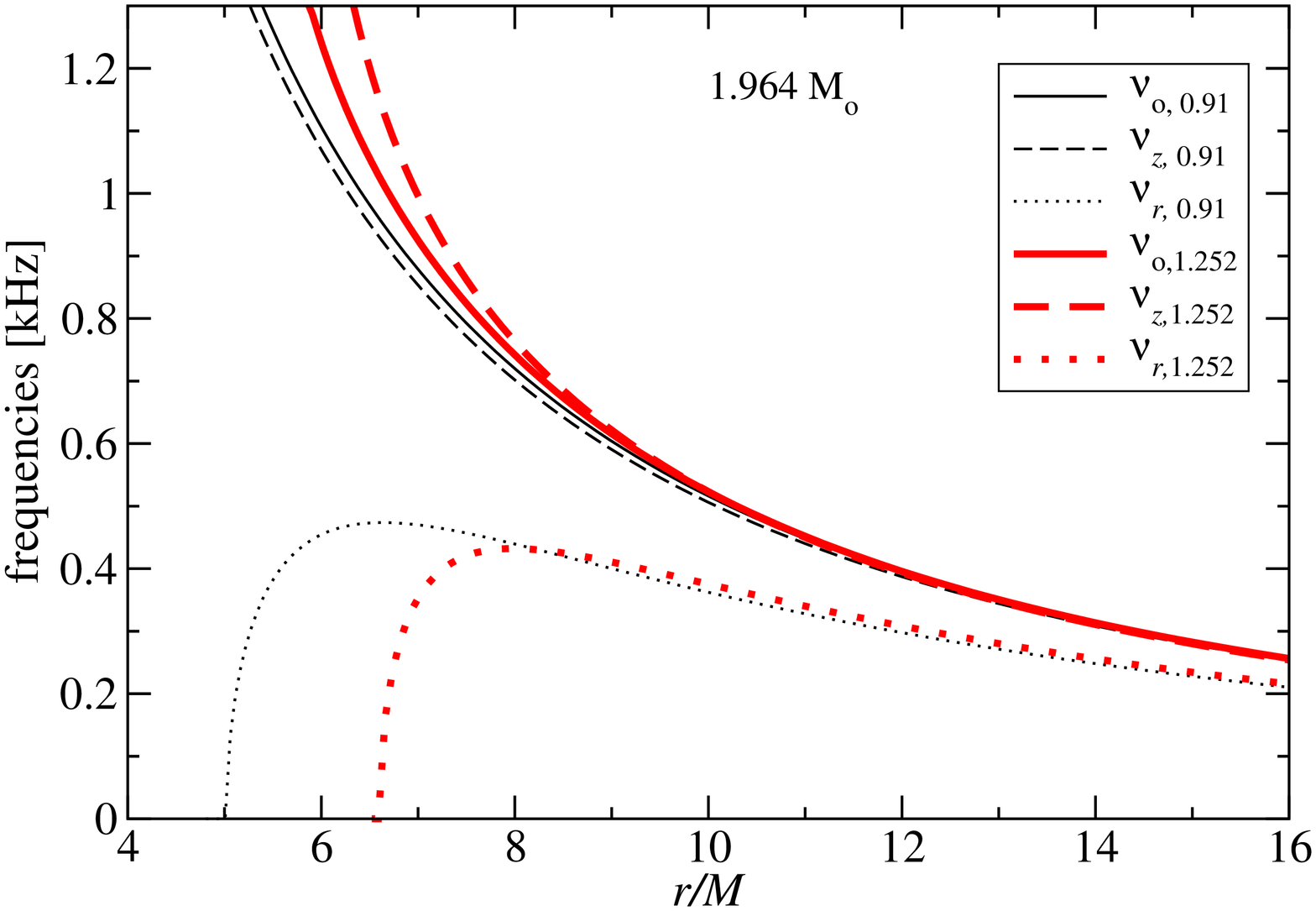}
\caption{Same as Fig.~\ref{velocity14}, but for $1.964 M_\odot$ at
  910  Hz (thin black lines) and 1252 Hz (thick red lines).}
\label{velocity19}
\end{center}
\end{figure}

We denote the orbital frequency
by $\nu_o$,  the vertical epicyclic frequency by $\nu_z$
and the radial epicyclic frequency by $\nu_r$. When comparing
them with the angular frequencies quoted in the literature,
one should bear in mind that the latter quantities are larger
by a factor of $2\pi$, e.g., the angular frequency of the radial
epicycle is $\kappa=2\pi\nu_r$.

The orbital and epicyclic frequencies of test particles in such a
metric is exhibited for two $M=1.4 M_\odot$ models in
Fig.~\ref{velocity14}.  The leftmost black curves correspond to a stellar
rotational frequency of 600 Hz. This is a modest rotation rate, so the
star may reasonably be described by the Hartle-Thorne \cite{Hartle68}
metric, which coincides through the first order in
stellar spin with the Kerr metric (see also \cite{Kluzn85}). As
expected, a marginally stable orbit is present slightly inwards of
$r=6M$ (the location where $\nu_r$, the black dotted curve in
Fig.~\ref{velocity14}, goes through zero),
 and the vertical epicyclic frequency is
slightly {\it lower} than the orbital frequency.\footnote{We suppress
  factors of $G/c^2$ in the text when discussing ratios such as
$r/M$, and $J/M^2\equiv a_*$.}
These results are consistent with those for a black hole with a small
value of the dimensionless Kerr parameter, $0<a_*<<1$ (e.g., \cite{Silbe01}).

The same quantities are shown in thick red lines
for a star of the same mass, but
spinning nearly twice as fast, at $1165\,$Hz.  Doubling the spin in
the Kerr metric would lead to a smaller ISCO radius, a higher maximum
value of the radial epicyclic frequency and a larger splitting between
the vertical epicyclic frequency and the orbital one, with
$\nu_z<\nu_o$. However, as is apparent from Fig.~\ref{velocity14},
doubling the rotational rate of the quark star leads to an increase of
the radius of the marginally stable orbit and a lowering of the
maximum value of $\nu_r$.  The splitting between the vertical
epicyclic and the orbital frequencies does indeed increase, but 
{\it the relative ordering} changes (for the radius range shown in the figure),
from $\nu_z<\nu_o$  at the 600 Hz rotation rate, to $\nu_z>\nu_o$ at
the 1165 Hz rate.
\begin{figure}{}
\begin{center}
\includegraphics[angle=-90,clip,width=0.52\textwidth]{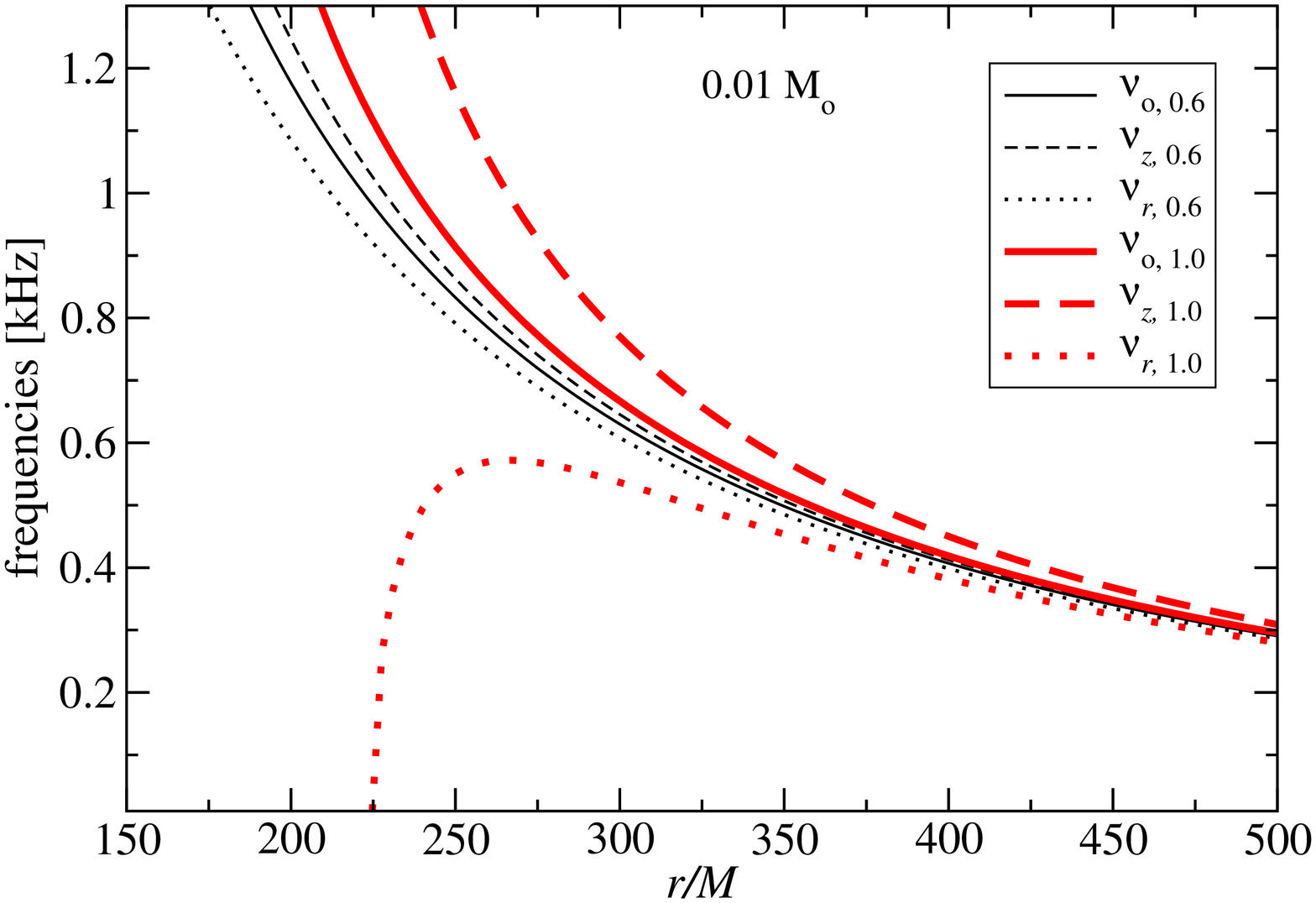}
\caption{Same as Fig.~\ref{velocity14}, but for $0.01 M_\odot$ at
  600 Hz (thin black lines) and 1165 Hz (thick red lines).}
\label{velocity01}
\end{center}
\end{figure}

\begin{figure}{}
\begin{center}
\includegraphics[angle=-90,clip,width=0.52\textwidth]{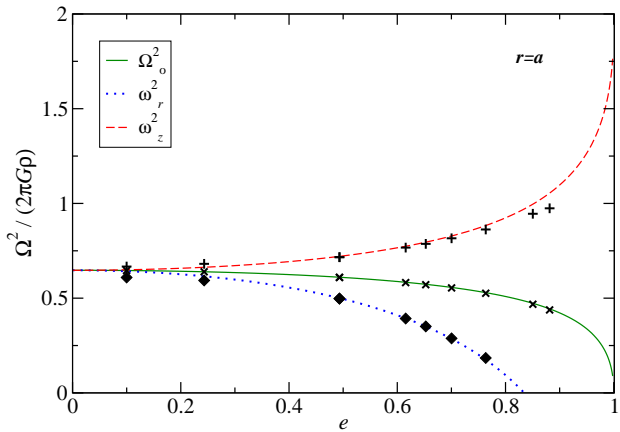}
\caption{Orbital and epicyclic frequencies (squared, in units of $2\pi
  G\rho$) at the stellar equator ($r=R_{\rm eq}\equiv a$) as a
  function of ellipticity of the Maclaurin spheroids (solid
  lines). Also shown (various symbols) numerical models of uniformly
  rotating strange quark stars of mass $M=0.001 M_\odot$ calculated by
  us with the general relativistic code RNS.  From top to bottom:
  vertical epicyclic frequency (red dashed line and + crosses),
  orbital frequency (green solid line and x crosses), radial epicyclic
  frequency (blue dotted line and diamonds).}
\label{velocity}
\end{center}
\end{figure}

\begin{figure*}
\begin{center}
\includegraphics[angle=-90,clip,width=0.49\textwidth]{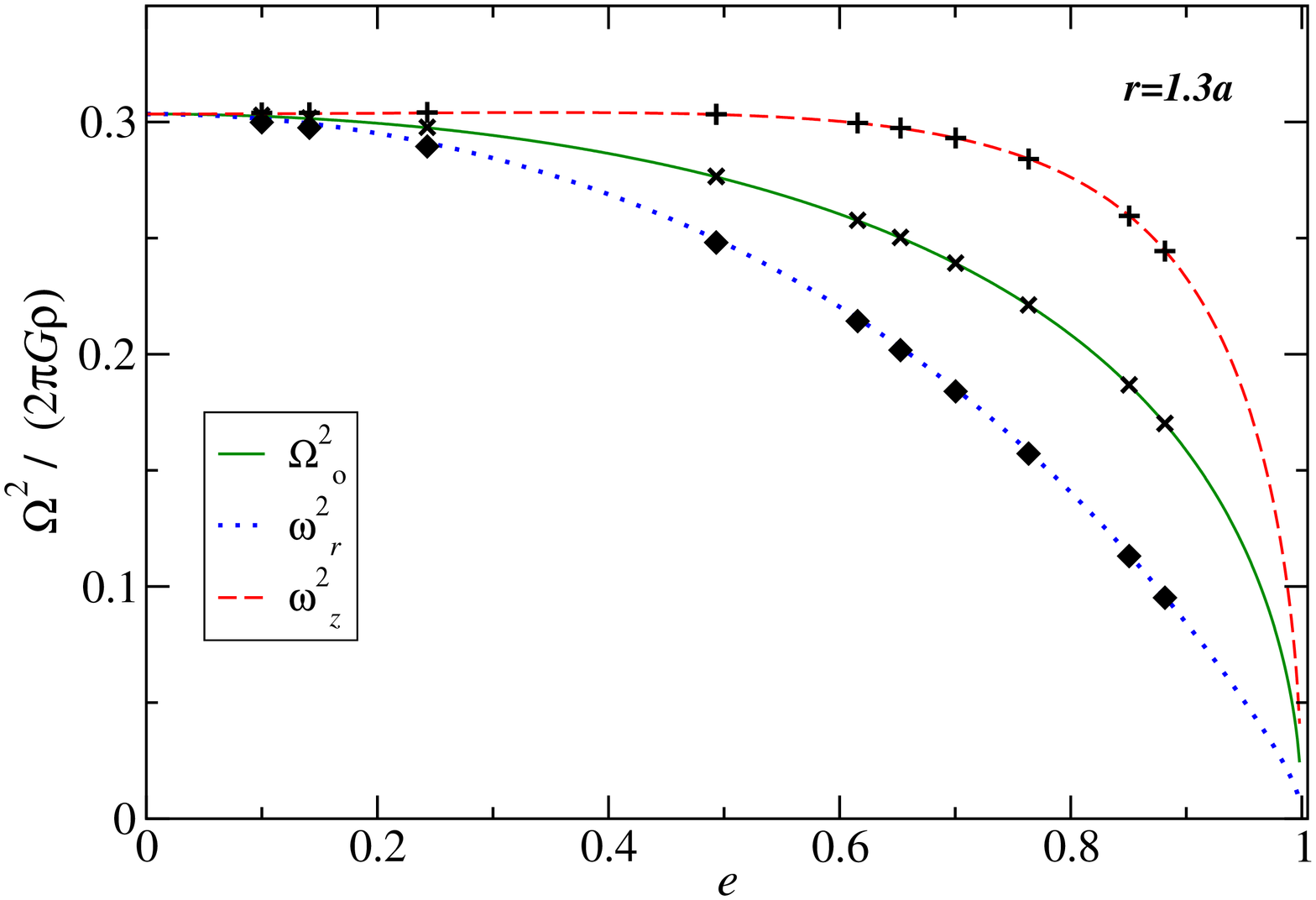}
\includegraphics[angle=-90,clip,width=0.49\textwidth]{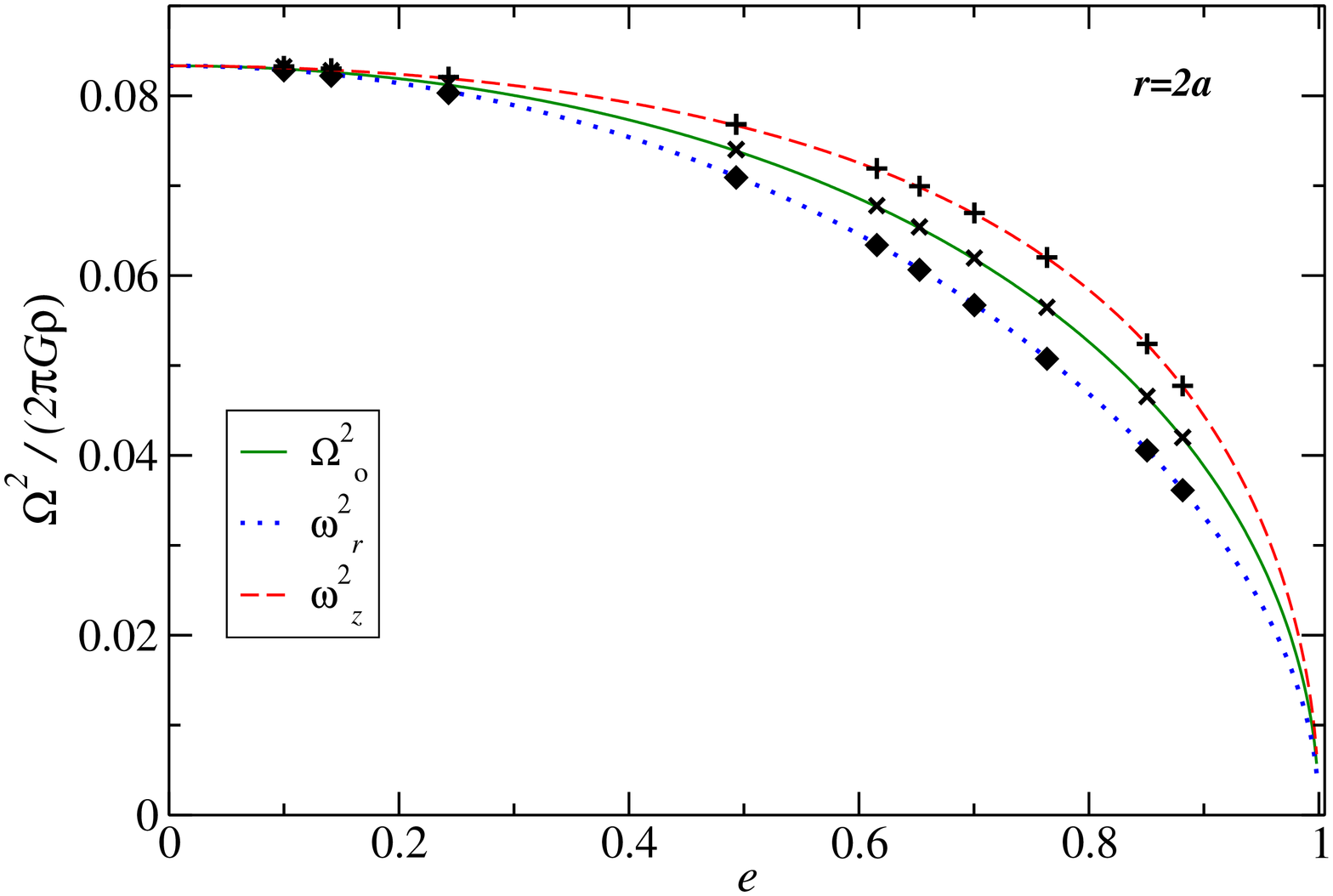}
\caption{Same as Fig.~\ref{velocity} but at $r=1.3a$ (left panel)
and  at  $r=2a$ (right panel).}
\label{velocityr2}
\end{center}
\end{figure*}

The same effects are exhibited by larger mass quark stars. 
Fig.~\ref{velocity19} shows the frequencies for $1.964 M_\odot$ models
rotating at 910 Hz (black curves) and at 1252 Hz (red thick curves).
Again, at the lower rotation rate the frequencies are qualitatively
similar to those of the Kerr metric, with the radius of the marginally
stable orbit $r_{\rm ms}<6M$, and $\nu_z<\nu_o$.
However, at the higher rotation rate,
$r_{\rm ms}>6M$ and $\nu_z>\nu_o$. 

Clearly, these departures
from a Kerr-metric behavior must be
related to the presence of
a rotationally distorted (``flattened'') figure of equilibrium
of the fluid,  i.e., a non-spherical mass distribution with
its higher gravitational multipoles.
To test this claim, we have computed the metric of a very low mass
quark star, of $M=0.01 M_\odot$. Such low masses are possible for
bodies composed of quark fluid, for unlike neutron star matter
which is unstable in the absence of strong gravity, quark matter
is self-bound (according to the Bodmer--Witten strange matter hypothesis
\cite{Bodmer71,Witten84}).

Fig.~\ref{velocity01} exhibits the frequencies of test particle orbits
for two $0.01 M_\odot$ models. For stars of such low mass, the orbits
essentially coincide with those in Newtonian gravity. For slowly rotating
fluid configurations, departures from spherical symmetry should remove
the degeneracy of the epicyclic frequencies and the orbital one that
is present in a $1/r$ potential, but all circular orbits should be
stable, i.e., the marginally stable orbit should be absent. This is
indeed the case for the 600 Hz model (shown in black), but notably not
true for a rapidly rotating model.  At 1000 Hz the radial epicyclic
frequency reaches a maximum at about $r\approx 270M$   and
goes to zero at $r_{\rm ms}=225M$. Note that these radii  are much larger than
the ones in Figs.~\ref{velocity14} and \ref{velocity19}, which are
comparable to the Schwarzschild values. Here, instead of being a low
multiple of the gravitational radius $M$, the marginally stable orbit radius
is comparable to the equatorial radius of the star $r_{\rm ms}\approx R_{\rm eq}$.
Clearly, at 1000 Hz the
rotational distortion of this low-mass star is sufficiently high for
the  Newtonian marginally stable orbit to appear.\footnote{In Newtonian
  gravity, the appearance of a marginally stable orbit requires a
  sizable octupole moment \cite{KluznBG01}.}
We note that regardless of the rotation rate, the ordering
of the orbital and the vertical epicyclic frequencies is
$\nu_r<\nu_0<\nu_z$ for this low mass model. As recently shown in
\cite{Gonde13}, this is the ordering of frequencies
in circular orbits in the Newtonian gravitational field of the classic
Maclaurin spheroids.  Therefore, we now turn to a discussion of the
Newtonian limit of rotating quark star models.

\section{Newtonian limit}
\label{limit}

\begin{figure*}
\begin{center}
\includegraphics[angle=-90,width=0.49\textwidth]{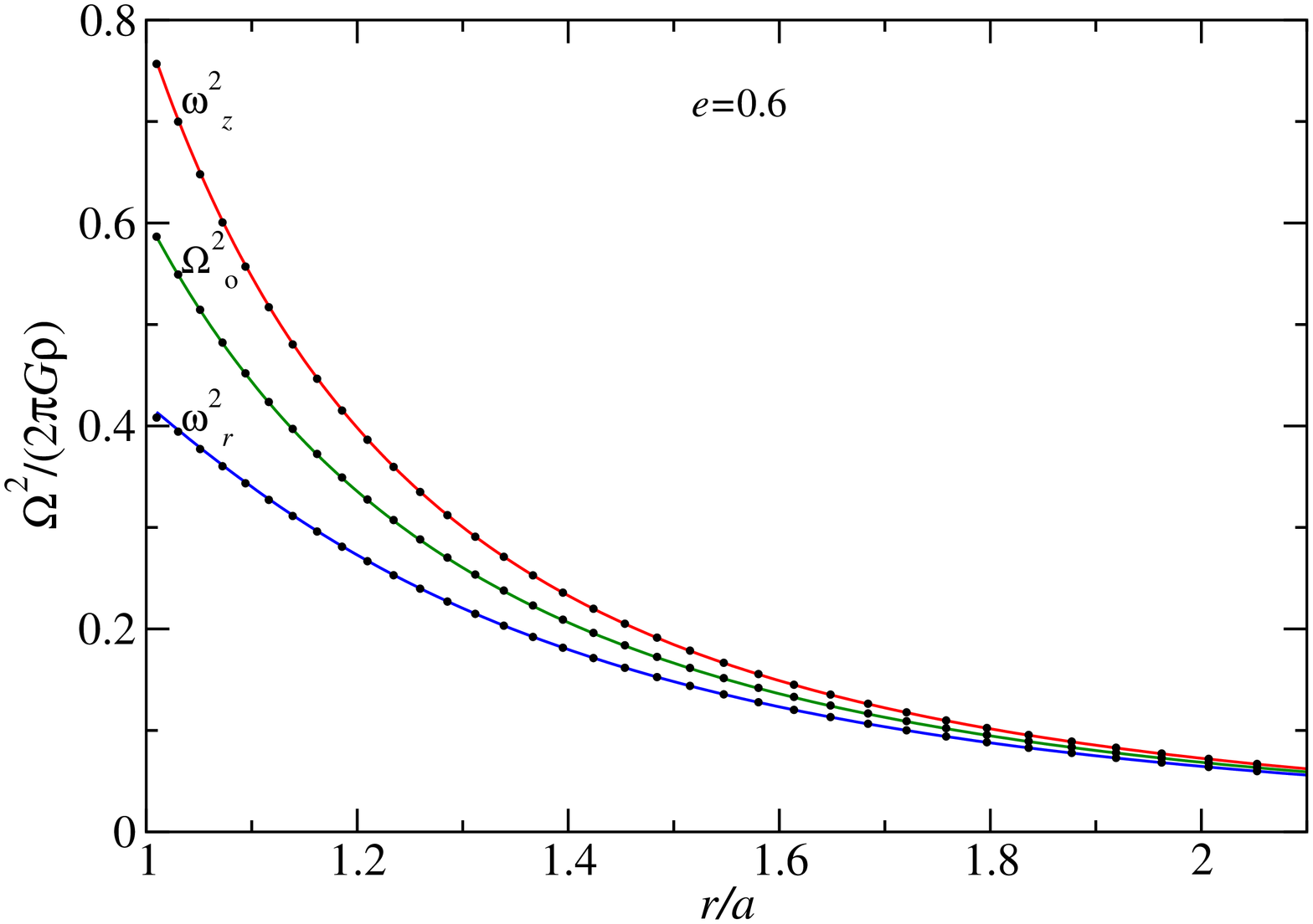}
\includegraphics[angle=-90,width=0.49\textwidth]{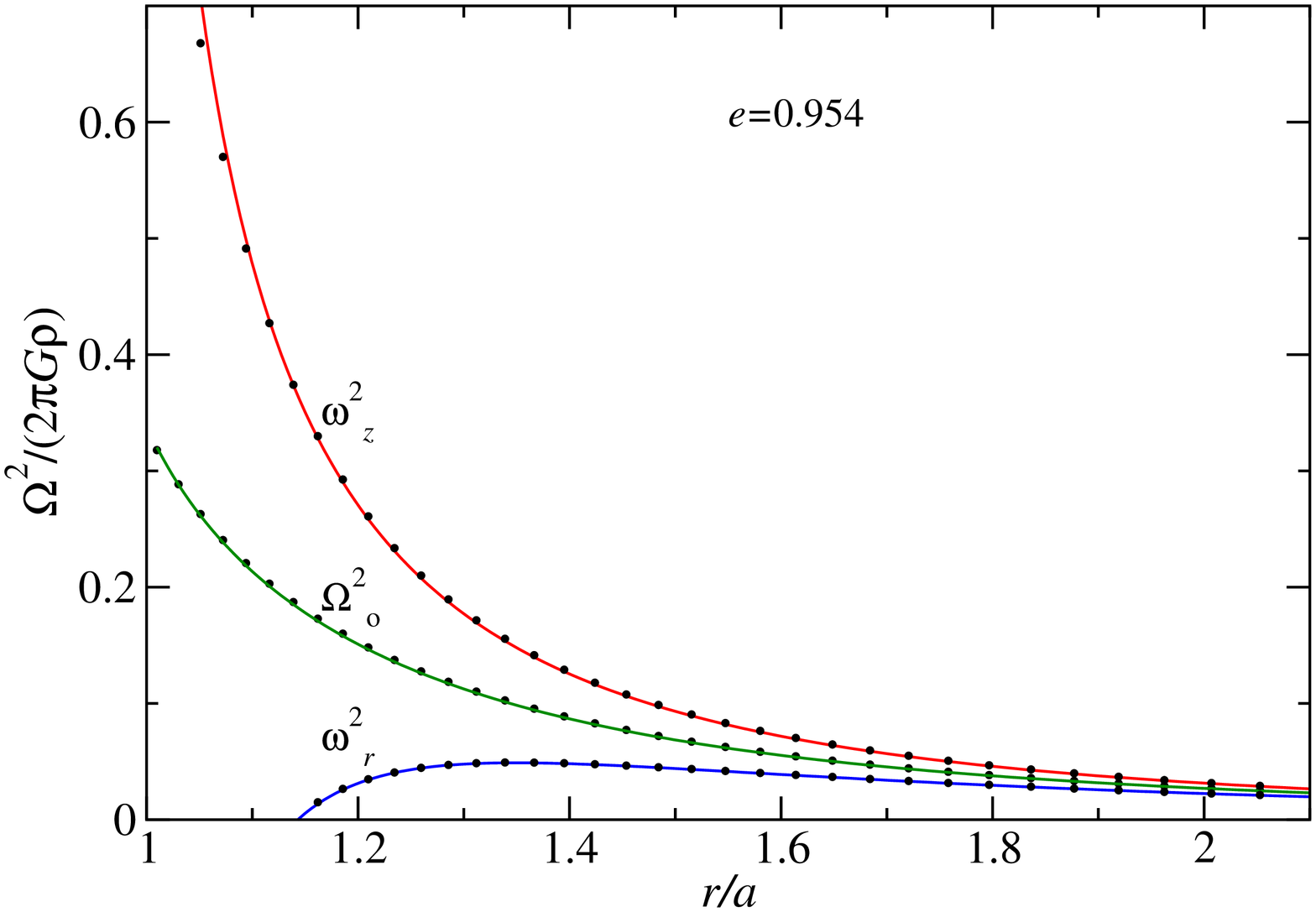}
\caption{Orbital and epicyclic frequencies (squared, in units of $2\pi
  G\rho$) versus radius (scaled with equatorial radius $a$) for
  Maclaurin spheroids (solid lines) and numerical models of  relativistic,
  uniformly rotating strange quark stars  of mass
  $M=0.001 M_\odot$, described by the MIT bag model
  (dots). For scaling purposes, the value of $\rho$ for SQS is 
  here assumed to be equal to the central density of the configuration.
  Left panel: stars with ellipticity $e=0.6$ (oblateness
  $0.2$). Right panel: $e=0.954$ (oblateness 0.7).
 From top to bottom:
  vertical epicyclic frequency (red), orbital frequency (green),
  radial epicyclic frequency (blue). }
\label{figmac}                                 
\end{center}                                
\end{figure*}

To discuss the Newtonian limit of our numerical results we compare
them with the known results for the classic figure of equilibrium of a
uniformly rotating fluid body,  with uniform density $\rho$.  
The analytic formulae for orbital and
epicyclic frequencies for test particles in motion around Maclaurin
spheroids can be found in ref. \cite{Gonde13}. We plot the
squares of these frequencies as a function of the ellipticity of the
spheroid in Figs.~\ref{velocity} and~\ref{velocityr2}, for three
different radii of the orbits, $r=a$, $r=1.3\,a$, and $r=2a$ ($a$
being the equatorial radius of the spheroid).  All frequencies scale
with the square root of the (uniform) density of the spheroid.
Overplotted with crosses and diamonds are the corresponding values
computed with our numerical code for a sequence of $M=0.001M_\odot$
quark star models.  The {\sl ellipticity}, $e$, is defined by 
$e^2=1-R_{\rm  p}^2/ R_{\rm eq}^2$, where $R_{\rm p}$ and $R_{\rm eq}$
are the  polar and equatorial radii of the spheroid, respectively.
We define {\sl oblateness} as  $1-R_{\rm  p}/ R_{\rm eq}$.
The overall agreement of our numerical results with the analytic formulae 
is excellent, the only significant 
discrepancies occurring for orbits grazing the equator (i.e., at $r=a$) 
at the highest ellipticities, and even then only for the vertical epicyclic 
frequency\footnote{ Since we are using relativistic SQS models, such
a level of agreement with uniform density Maclaurin spheroids is still
remarkable and the slight discrepancy in the vertical epicyclic frequency
at the surface of the star indicates that this particular frequency is 
sensitive to the precise multipolar structure of the gravitational field.}.

We continue the comparison in Fig.~\ref{figmac}, now plotting the
scaled frequencies squared as a function of the radius (in units of
the equatorial radius) at fixed ellipticities of $e=0.6$ and
$e=0.954$. Again, there is excellent agreement between the numerical
results obtained with the RNS code for relativistic quark star
models and the analytic formulae for Maclaurin spheroids. Note the
presence of the marginally stable orbit ($\nu_r^2=0$) outside the
stellar surface for $e=0.954$, but not for $e=0.6$. This is in
agreement with Fig.~\ref{velocity} where, clearly, $\nu_r^2<0$ for $e=0.954$ at
$r=a$, and $\nu_r^2>0$ for $e=0.6$ at $r=a$, implying instability of
circular orbits at the equator in the former case and stability of
orbits at the equator in the latter case.

As remarked in \cite{Gonde13}, the overall relationship
of the epicyclic and orbital frequencies is reminiscent of retrograde
orbits in the Kerr geometry. However, in the Newtonian limit represented
in Fig.~\ref{figmac}, there is
no distinction between prograde and retrograde orbits.


\section{Astrophysical implications}
\label{astro}

We have found that rotationally induced oblateness strongly affects
the epicyclic frequencies in relativistic stars. It has already been
noted in the literature that the position of the marginally stable
orbit is pushed out by rapid stellar rotation \cite{Sterg99}.
We now note that this
is related to a general decrease of the radial epicyclic frequency.
In contrast, the vertical epicyclic frequency increases with the
oblateness of the star, thus the difference between
the two epicyclic frequencies increases with the stellar rotation rate.
These effects may have interesting astrophysical implications,
particularly if it turns out that similar qualitative effects
obtain for stars modelled with the conventional neutron star matter
equations of state (this is a subject of an ongoing investigation).

 High frequency QPOs (HFQPOs) are variations in the light curves of
 accreting neutron stars and black holes \cite{Klis00}.  In LMXBs,
 because of low photon counts (i.e., small area of currently existing
 detectors) they are currently observed only in the Fourier transform
 of the light curves, as peaks of finite width in the power density
 spectra\footnote{HFQPOs seem to have their counterparts in two
   supermassive black holes, where they can be observed directly in
   the light curve. These are Sgr A* at the centre of the Galaxy, with
   its 17-minute QPO, and
   NGC 5408 X-1 \cite{Strohm07}.}. Their quality factor varies from a
 few in the black hole case, to $Q>100$ in neutron stars
 \cite{Barret05}.  Very often the HFQPOs occur in pairs, in neutron
 star sources the two frequencies correlate in a reproducible fashion
 with other properties of the source (such as luminosity).

Nearly all models of HFQPOs
in neutron stars and black holes
involve some combination of the epicyclic frequencies and the orbital frequency.
Therefore, a marked change of the epicyclic frequencies related to the
stellar oblateness will affect the predictions of the models, as well as
the differences between HFQPOs in the neutron/quark
star and black hole systems (if the models are correct).

In the relativistic precession model (RPM) \cite{Stella99}, the upper
kHz QPO is taken to be the orbital frequency, while the lower one the
difference between the orbital frequency and the radial epicyclic
frequency. Applying this model to our results for quark stars, one
would expect that the lower kHz QPO should increase in frequency
with the rotation
rate of the star. This in itself would not be a dramatic effect.
However, in general, if at least one frequency involves the orbital
frequency, one would expect a cut-off at the marginally stable orbit
\cite{Kluzn90}, and since the stellar mass is usually inferred in such
a case from the highest observed frequency, which is taken to be the orbital
frequency in the marginally stable orbit
\cite{Kluzn90,Kluzn98,Kaaret99}, $M$ would have been overestimated if
in fact the marginally stable orbit is pushed out by rapid stellar
rotation, as we find for the SQS models.

\begin{figure}{}
\includegraphics[angle=-90,width=0.52\textwidth]{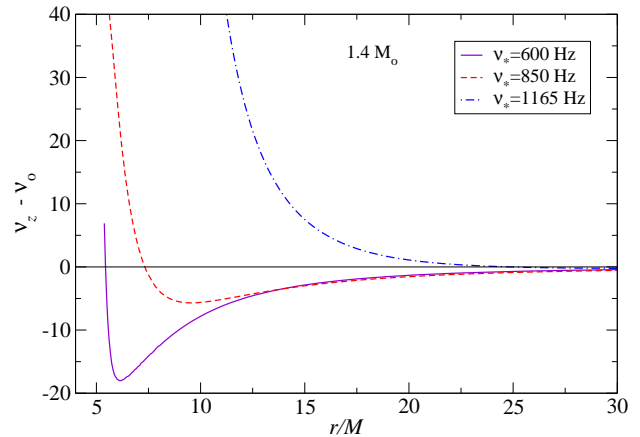}
\vskip 0.5cm

\caption{The difference between vertical epicyclic frequency and
  orbital frequency for a $1.4 M_\odot$ rigidly rotating strange quark
  star rotating with frequency 1165 Hz (blue dot-dashed line), 850 Hz
  (red dashed line) and 600 Hz (violet solid line). }
\label{f:difference}
\end{figure}

In diskoseismology \cite{Kato01,Nowak92}, the lower kHz QPO is
identified with the {\sl g}-mode \cite{Kato80} trapped near the
maximum of the radial epicyclic frequency, while the higher kHz QPO
with a {\sl c}-mode \cite{Silbe01}. For rapidly rotating quark stars,
the frequency of the lower kHz QPO would then be diminished in this
model (in contrast to the RPM). Perhaps the most prominent effects
would be on the upper kHz QPO if it corresponds to a 
{\sl trapped  c-}mode. It has been demonstrated \cite{Silbe01}
that no trapped {\sl c}-modes
exist if $\nu_z-\nu_o>0$ is a decreasing function with $r$, as is
indeed  the case close to the rotating quark star
(Fig.~\ref{f:difference}).  In this model it seems likely that that
the higher kHz QPO would be absent in  the emissons of the inner
accretion disk of rapidly rotating quark stars. If the  {\sl  c}-modes
are absent down to the minimum of $\nu_z-\nu_0$,  
Figs.~\ref{f:difference}, \ref{velocity14} suggest that already for
a stellar rotation rate lower than 900 Hz, the upper kHz QPO could be lower
by about 300 Hz than for slowly rotating stars (because the {\sl  c}-mode
would truncate at $r\approx 9M $ instead of $r\approx 6M$).
However, the status of
 {\sl  c}-modes in the case when $\nu_o-\nu_z>0$ is an increasing function
of $r$ is unclear at present. The carefully investigated case
\cite{Silbe01} is that of the Kerr metric, where  $\nu_z-\nu_o<0$ and
$\nu_0-\nu_z$ is a decreasing function of $r$ for an accretion disk
co-rotating with the black hole.  More definitive conclusions must
await a consensus on the origin of kHz QPOs.

\section{Discussion}
\label{disc}

In general relativity, the properties of the marginally stable orbit
 are familiar from work in the Kerr metric \cite{Bardeen72}.  For
prograde orbits the radius of the ISCO, in units of the gravitational
mass, is a monotonically decreasing function of the black hole spin,
$a_*$.  For retrograde orbits, it is a monotonically increasing
function of $a_*$.  In the Schwarzschild metric the vertical epicyclic
frequency is equal to the orbital frequency. In the Kerr metric, the
vertical epicyclic frequency is lower than the orbital frequency for
prograde orbits and larger for retrograde orbits (e.g., \cite{Silbe01}).  
The lower the value of the ISCO radius, the higher the
maximum value of the epicyclic frequency. 

We found that models of rapidly rotating SQS 
of astrophysically relevant masses ($M_\odot<M<2M_\odot$) have epicyclic
frequencies whose behavior differs qualitatively from that of prograde
orbits in the Kerr geometry. The marginally stable orbit is pushed away
from the star to values $r/M>6$, while near the star the vertical epicyclic
frequency is larger than the orbital frequency
(Figs.~\ref{velocity14},~\ref{velocity19}).

Qualitatively similar statements hold for rotating neutron stars,
although the effect of rotation is less pronounced \cite{Kluzn04}.
Stergioulas, Klu\'zniak, Bulik \cite{Sterg99} computed the location of
the ISCO for rapidly rotating massive quark stars and noted that
rotation-induced  oblateness pushes the ISCO out to larger radii. We
are now {\bf explaining these effects by considering the epicyclic
frequencies of circular orbits around rotating SQS and we extended our
numerical investigation to low-mass models to compare our results with
the Newtonian theory of rapidly rotating figures of equilibrium.}
The orbital and epicyclic frequencies are
of great interest in the context of X-ray observation of
high-frequency variability of emission, specifically of the kHz QPOs
\cite{Kato80,Kluzn90,Nowak91,Stella99,Klis00,Abram01,Kato01,Kluzn05}.

Our relativistic numerical calculations of the orbital frequencies around
low-mass stars, 
are found to agree well with the analytical calculations for Maclaurin
spheroids, given in \cite{Amste02} and \cite{Gonde13}. 
In the Newtonian limit, the gap between the marginally stable
orbit and the stellar surface is produced by the oblateness of the
rapidly rotating low-mass quark stars
 \cite{KluznBG01,Zduni01,Amste02}.

\section{Conclusions}
We have computed numerical models of rapidly rotating strange quark stars
and their external metric. In particular, we have computed the orbital
and epicyclic
frequencies of circular prograde orbits around SQS
for astrophysically relevant stellar masses, i.e., those occurring in LMXBs.

{\bf We have validated the epicyclic frequency module of the RNS code
by comparing the code results for low-mass quark star models,
at  $M=0.01 M_\odot$ and $M=0.001 M_\odot$, with analytical formulae.}

We find that the properties of orbital and epicyclic frequencies
are a result of the interplay of competing GR and Newtonian effects.
For moderately rotating massive stars the
behavior of the epicyclic frequencies is very similar to that of the
frequencies in neutron stars, which in turn are similar to those of
prograde orbits in the Kerr metric of slowly spinning black holes
($0<a_*<<1$). However, at high rotation rates the behavior of the
epicyclic frequencies near the quark star is similar to that of
retrograde orbits in the Kerr metric, i.e., in the latter case the
marginally stable orbit is pushed away from the star and the vertical
epicyclic frequency is higher than the orbital one.

For moderately rotating massive SQS the behavior
of the epicyclic frequencies is dominated by GR effects.  However, for
rapidly rotating SQS a qualitatively new effect appears for
prograde orbits---the vertical epicyclic frequency becomes larger than
the orbital frequency. This is a non-relativistic effect of
oblateness, known from a study of Maclaurin spheroids \cite{Gonde13},
as has been verified in a calculation of the
frequencies of low mass stars, for which the effects of GR are
unimportant, and for which it is found
the vertical epicyclic frequency is always larger than
the orbital frequency, even at very low stellar rotation rates
(Figs.~\ref{velocity}--\ref{figmac}).

The competition of GR effects and those of higher multipoles can clearly
be seen in a plot of the difference between the vertical epicyclic frequency
and the orbital one. Fig.~\ref{f:difference} shows that for a $1.4 M_\odot$
star, at large radii the vertical epicyclic frequency is lower than the orbital
one, even at a rotation rate as high as 1165 Hz. It is only close to the star
that the effect of higher multipoles prevails and raises the value of
the vertical epicyclic frequency above the orbital one.
Were this effect, affecting
the kHz QPOs (\S~\ref{astro}), to occur also in neutron stars, it could have
observable implications in many LMXBs.

\acknowledgments 
We thank the anonymous referee for remarks which helped us to
improve the presentation. 
Discussions with professor R.V. Wagoner about diskoseismology are gratefully
acknowledged.
This work was partially supported by Polish grants NN203 511 238,
2013/08/A/ST9/00795, POMOST/2012-6/11 Program of Foundation for Polish
Science, 2013/01/ASPERA/ST9/00001 and by CompStar -- a Research
Networking Programme of the European Science Foundation.



\begin{thebibliography}{}
\bibitem{Weber99}
F. Weber, JPhG, {\bf 25}, 195 (1999).

\bibitem{Bodmer71}
A. R. Bodmer, Phys. Rev. D {\bf 4}, 1601 (1971).

\bibitem{Witten84}
E. Witten, 
\newblock Phys. Rev. D {\bf 30}, 272 (1984).

\bibitem{Alcock86}
C.~Alcock, E.~Farhi, and A. Olinto,
Astrophys. J. {\bf  310}, 261 (1986).

\bibitem{Haens86}
P. Haensel, J. L.  Zdunik, R. Schaeffer,
Astron. Astrophys.  {\bf 160}, 121 (1986).

\bibitem{Fahri84}
E. Fahri, and R. L. Jaffe, 
\newblock Phys. Rev. D  {\bf 30}, 2379 (1984).

\bibitem{Gonde03}
D. Gondek-Rosi\'nska, E. Gourgoulhon, P. Haensel, 
\newblock Astron. Astrophys.  {\bf 412}, 777 (2003).

\bibitem{Kapla49} 
S.~A. Kaplan, ZhETF, {\bf 19}, 951 (1949).

\bibitem{Shaku73}
N. I. Shakura, R. A. Sunyaev,  Astron. Astroph. {\bf 24}, 337 (1973);
I. D. Novikov, K. S. Thorne,  in Black Holes (Les Astres Occlus), ed.
C. DeWitt, B. S. DeWitt (Gordon and Breach Science Publishers), 343 (1973).

\bibitem{Sadow11}
A. S{\c a}dowski et al.,
 Astron.  Astroph. {\bf 532}, A41 (2011).

\bibitem{Kluzn85}
W. Klu{\'z}niak, R. V. Wagoner, Astrophys. J. {\bf 297}, 548 (1985).

\bibitem{Cook94}
G. B. Cook, S. L. Shapiro, S. A. Teukolsky, Astrophys. J {\bf 422}, 227 (1994).

\bibitem{Kluzn90}
W. Klu{\'z}niak, P. Michelson, R. V.  Wagoner,
 Astrophys.~J. {\bf 358}, 538 (1990);
W. Klu{\'z}niak, S. Rappaport, Astrophys.~J. {\bf 671}, 1990 (2007);
M. Bejger,  M. Fortin, P.~Haensel, J. L. Zdunik,
  Astron. Astroph. {\bf 536A}, 87 (2011).

\bibitem{Gonde01}
D. Gondek-Rosi\'nska, N. Stergioulas, T. Bulik, W. Klu\'zniak,
E. Gourgoulhon, Astron. Astrophys. {\bf 380}, 190 (2001).

\bibitem{Bardeen72}
J. M. Bardeen, W. H. Press, S. A. Teukolsky,
 Astrophys.~J. {\bf 178}, 347 (1972).

\bibitem{Pugli11}
D. Pugliese, H. Quevedo, R. Rufini, Phys. Rev. D {\bf 83},
024021 (2011).

\bibitem{Vieira13}
R. S. S. Vieira, et al., preprint arXiv1311.5820

\bibitem{Kato80}
S. Kato, J. Fukue, PASJ {\bf 32}, 377 (1980).

\bibitem{Nowak92}
M. Nowak, R. V. Wagoner,   Astrophys. J. {\bf 393}, 697 (1992);
C. A. Perez, A. S. Silbergleit, R. V. Wagoner, D. E. Lehr,
Astrophys. J. {\bf 476}, 589 (1997).

\bibitem{Silbe01}
A. S. Silbergleit, R. V. Wagoner, M. Ortega-Rodr\'iguez,
 Astrophys. J. {\bf 548}, 335 (2001).

\bibitem{Klis00}
M. van der Klis, ARA\&A  {\bf 38}, 717 (2000).

\bibitem{KluznBG01}
W. Klu{\'z}niak W., T. Bulik, D.  Gondek-Rosi{\'n}ska,
 ESA-SP {\bf 459}, 301 [astro-ph/0011517] (2001).
 
\bibitem{Zduni01}
J. L. Zdunik, E. Gourgoulhon, Phys. Rev. D {\bf 63}, 087501 (2001).

\bibitem{Amste02}
Amsterdamski P., Bulik T., Gondek-Rosi\'nska D., Klu\'zniak W.,
 Astron. Astrophys. {\bf 381}, L21 (2002).

\bibitem{StergF95}
N. Stergioulas, J. L. Friedman, Astrophys. J. {\bf 444}, 306 (1995).

\bibitem{Kluzn98}
W. Klu{\'z}niak, Astrophys. J. {\bf 509}, L37 (1998).

\bibitem{Morsink99}
S. M. Morsink, L. Stella,  Astrophys. J. {\bf 513}, 827 (1999).

\bibitem{Hartle68}
J. B. Hartle, K. S.  Thorne, Astrophys. J. {\bf 153}, 807 (1968).

\bibitem{Gonde13}
W.  Klu\'zniak, D. Rosi\'nska, MNRAS {\bf 434}, 2825 (2013).

\bibitem{Lorene}
http://www.lorene.obspm.fr/

\bibitem{Madsen99}
J. Madsen, Phys. Rev. Lett. {\bf 81} 3311 (1998);
J. Madsen, in {\sl Hadrons in Dense Matter and Hadrosynthesis}, Springer,
p. 162 (1999);
J. E. Horvath, Int. J. Mod. Phys. D, {\bf 8}, 669 (1999);
E. Gourgoulhon, Astron. Astrophys.  {\bf 349}, 851 (1999);

\bibitem{Sterg99}
N. Stergioulas, W. Klu\'zniak, T. Bulik,
 Astron. Astrophys.  {\bf 352}, L116 (1999).

\bibitem{ZduniBKHG00}
J. L. Zdunik, T. Bulik, W. Klu\'zniak, P. Haensel, D. Gondek-Rosi\'nska,
 Astron. Astrophys.  {\bf 359}, 143 (2000).

\bibitem{Sterg03}
N. Stergioulas, Living Rev. Relativity  {\bf 6}, 3,
http://www.livingreviews.org/lrr-2003-3 (2003).

\bibitem{Kamat89}
H. Kamatsu, Y. Eriguchi, I. Hachisu,
 Mon. Not. R. Astron. Soc. {\bf 237}, 355 (1989).

\bibitem{Kluzn04}
W. Klu\'zniak, M. A. Abramowicz, S. Kato, W. H.  Lee, N.  Stergioulas,
 Astrophys. J. {\bf 603}, L89 (2004).

\bibitem{Marko00}
D. Markovic, F. K.  Lamb, arXiV:astro-ph/0009169 (2000).

\bibitem{Strohm07}
T. E. Strohmayer, R. F. Mushotzky, L. Winter, R. Soria, P. Uttley, M. Cropper,
 Astrophys. J. {\bf 660}, 580 (2007).

\bibitem{Barret05}
D. Barret, W. Klu\'zniak, J. F. Olive, S. Paltani, G. K. Skinner,
Mon. Not. R. Astron. Soc. {\bf 357}, 1288 (2005).

\bibitem{Stella99}
L. Stella, M. Vietri, S. Morsink, ApL\&C, 38, 57 (1999).

\bibitem{Kaaret99}
P. Kaaret, S. Piraino, P. F. Bloser, E. C. Ford, J. E. Grindlay,
A. Santangelo, A. P. Smale, W. Zhang, Astrophys. J. {\bf 520}, L37

\bibitem{Kato01}
R. V. Wagoner, Phys. Rep. {\bf 311}, 259 (1999);
S. Kato, PASJ {\bf 53}, 1 (2001).

\bibitem{Nowak91}
M. A. Nowak, R. V. Wagoner, Astrophys. J. {\bf 378}, 656 (1991).

\bibitem{Abram01}
M. A. Abramowicz, W. Klu{\'z}niak, Astron. Astrophys. {\bf 374}, L19 (2001).

\bibitem{Kluzn05}
W. Klu{\'z}niak, AN, {\bf 326}, 820 (2005);
G. T\"or\"ok, M. A. Abramowicz, W.  Klu\'zniak, Z. Stuchl\'ik,
Astron. Astrophys. {\bf 436}, 1 (2005).


\end{thebibliography}
\end{document}